# Future large-scale water-Cherenkov detector

L. Agostino, M. Buizza-Avanzini, M. Marafini, T. Patzak, and A. Tonazzo[*]

*APC, Université Paris Diderot, CNRS/IN2P3, CEA/Irfu, Observatoire de Paris, Sorbonne Paris Cité, F-75205 Paris Cedex 13, France*

M. Dracos and N. Vassilopoulos

*IPHC, Université de Strasbourg, CNRS/IN2P3, F-67037 Strasbourg, France*

D. Duchesneau

*LAPP, Université de Savoie, CNRS/IN2P3, F-74941 Annecy-le-Vieux, France*

M. Mezzetto

*INFN Sezione di Padova, 35131 Padova, Italy*

L. Mosca

*Laboratoire Souterrain de Modane, F-73500 Modane, France*

MEMPHYS (MEgaton Mass PHYSics) is a proposed large-scale water-Cherenkov experiment to be performed deep underground. It is dedicated to nucleon decay searches and the detection of neutrinos from supernovae, solar, and atmospheric neutrinos, as well as neutrinos from a future beam to measure the $CP$ violating phase in the leptonic sector and the mass hierarchy. This paper provides an overview of the latest studies on the expected performance of MEMPHYS in view of detailed estimates of its physics reach, mainly concerning neutrino beams.

## I. INTRODUCTION

A megaton-scale water-Cherenkov detector would have competitive capabilities for accelerator-based neutrino oscillation physics. In addition, it would reach a sensitivity on the proton lifetime close to the predictions of most supersymmetric or higher dimension grand unified theories and it would explore neutrinos from supernovae and from other astrophysical sources.

Such a detector is most attractive because it relies on a well established technique, already used by the IMB [1], KamiokaNDE [2], and SuperKamiokande (SK) [3] experiments. It would be roughly 10 times the size of SK, a reasonable extension of a known, well performing detector.

An expression of interest for such a project, called MEMPHYS (MEgaton Mass PHYSics), has been prepared [4].

The potential for neutrino physics with specific neutrino beams was investigated in detail in [5]. The authors assumed the same performance as the SK detector in terms of detection efficiency, particle identification capabilities, and background rejection. The behavior of a larger scale detector will, however, be different, because of the larger distance traveled by light to reach the photomultipliers.

The physics panorama has also changed since then, with the measurement of the $\theta_{13}$ mixing angle. While old analyses were optimized for values of $\sin^2(2\theta_{13})$ of the order of 0.001 or 0.01, the measured value has been found just underneath the CHOOZ limit, of the order of 0.1 [6–8]. As a consequence, the amplitude of the leptonic $CP$ violation (LCPV) term, for each given value of the $CP$ violating phase $\delta_{CP}$, will be smaller than assumed before. This requires some revisiting of the previous analyses and considerations. It is thus essential to estimate the statistical error in the LCPV measurement with a given fiducial mass, beam intensity, and time of exposure and compare it with the corresponding systematic errors, once the beam has been optimized. Only then will it be possible to define the best strategy to be adopted for the LCPV measurement. This has been one of the main topics developed in the context of the EUROnu EU-FP7 Design Study [9].

Two types of possible future neutrino beams are considered for oscillation measurements with a water-Cherenkov detector: (i) "Super-Beams" [10] are high-intensity neutrino beams generated with a "traditional" technique: pions and kaons, produced in accelerated proton collisions on a target, decay in flight giving origin to beams consisting mainly of $\nu_\mu$ or $\bar{\nu}_\mu$, with a contamination from other neutrino species at the percent level and not precisely known; (ii) "Beta-Beams" [11,12] are made of neutrinos

---

originated from the decay in flight of accelerated beta emitters (such as $^{18}$Ne or $^{6}$He) and consist purely of $\nu_e$ or $\bar{\nu}_e$ with a known energy spectrum.

Optimization and cost estimates for these beams have been studied in EUROnu [13].

In this paper, a realistic evaluation of the expected MEMPHYS performance is presented: after a description of the detector layout in Sec. II and of some R&D towards its construction in Sec. III, the full simulation of the detector is presented in Sec. IV and its use for the optimization of the design is shown in Sec. V. Realistic analysis algorithms have been developed and implemented on the simulation, as reported in Secs. VI and VII, and their performance is summarized in terms of "migration matrices" from true to reconstructed neutrino energy, which can be used to evaluate the physics reach, as illustrated in Sec. VIII. The impact of systematic uncertainties on the physics potential of the detector is discussed in Sec. IX.

## II. THE MEMPHYS DETECTOR

MEMPHYS is a proposed large-scale water-Cherenkov detector with a fiducial mass of the order of half a megaton.

The detector could be installed at the Fréjus site, near the existing *Laboratoire Souterrain de Modane* (LSM), in the tunnel connecting France to Italy, located at 130 km from CERN and with a rock overburden of 4800 meters water equivalent. Possible installation at other European sites was studied in the context of the LAGUNA EU-FP7 Design Study [14].

The original plan [4] envisaged three cylindrical detector modules of 65 meters in diameter and 60 meters in height. At the Fréjus site, the characteristics of the rock allow for a larger excavation in the vertical direction. With an updated design of 80 m height, for example, the fiducial volume would be increased to 572 kilotons (30% larger). Heights up to 103 m are possible, which would allow for the same total fiducial mass with only two modules. The new reference design [15] envisages two modules of 103 m height and 65 m diameter. Taking into account a 1.5 m thick veto volume surrounding the main tank and a cut at 2 m from the inner tank wall for the definition of the fiducial volume, as done in SK to allow for Cherenkov cone development, the total fiducial mass would be 500 kilotons.

Each module is equipped with ~120000 8" or 10" photomultipliers (PMTs) providing 30% optical coverage (equivalent, in terms of number of collected photoelectrons, to the 40% coverage with 20" PMTs of SK).

A small amount of gadolinium salt in the water [16] would enhance the detector capabilities with identification of the neutron from inverse-$\beta$ $\bar{\nu}_e$ interactions, thus reducing the background and allowing to lower the threshold for physics studies in this channel. No detailed study has been performed for this paper on this item, which is not relevant for beam neutrinos.

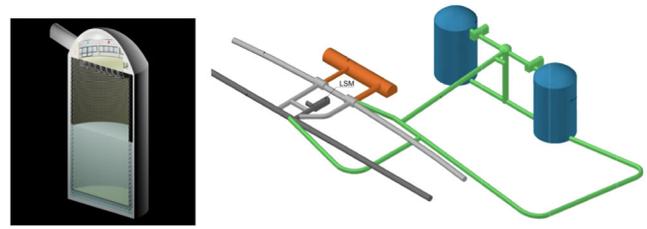

FIG. 1. Schematic view of one MEMPHYS module (left) and design for installation and infrastructure at a possible extension of the LSM underground laboratory at the Fréjus site (right) (courtesy of Lombardi Engineering S.A.). Each tank is 65 m in diameter and 103 m in height. The total fiducial mass is 500 kton.

A schematic view of the detector and of a possible layout for installation at the Fréjus site are shown in Fig. 1 (courtesy of Lombardi Engineering S.A.).

## III. R&D TOWARDS LARGE LIQUID-BASED DETECTORS

One important R&D item towards the construction of MEMPHYS and other large liquid-based detectors, such as LENA [17] or GLACIER [18], is focused on the reduction of the number of electronics channels for power supply and signal readout of the PMTs, which is expected to be one of the major costs of the experiment.

The PMm$^2$ R&D program [19] has developed an integrated readout electronics circuit (an ASIC called PARISROC [20]) for groups of PMTs (matrix of $4 \times 4$).

The electronics and acquisition are being fully tested with the MEMPHYNO prototype [21], a test bench for light sensor or electronics solution for next generation large size experiments. MEMPHYNO is installed at the APC Laboratory in Paris. It consists of a high-density polyethylene tank of $2 \times 2 \times 2$ m$^3$, filled with water and placed inside a muon hodoscope. The hodoscope is composed of four planes of plastic scintillator bars (kindly provided by the OPERA collaboration), two above and two below the tank, with X-Y measurement of the track coordinates. It provides the trigger for the passage of a muon and a reconstruction of the track direction.

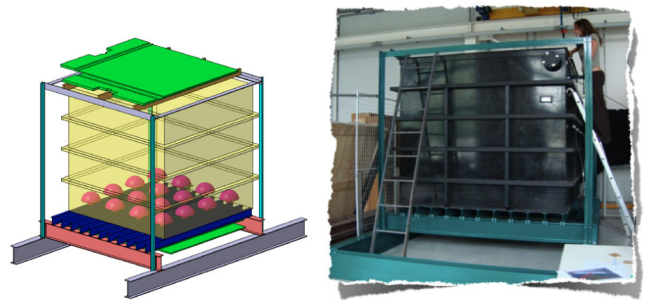

FIG. 2. Schematic view of the MEMPHYNO test bench (left) and photograph of the facility (right).

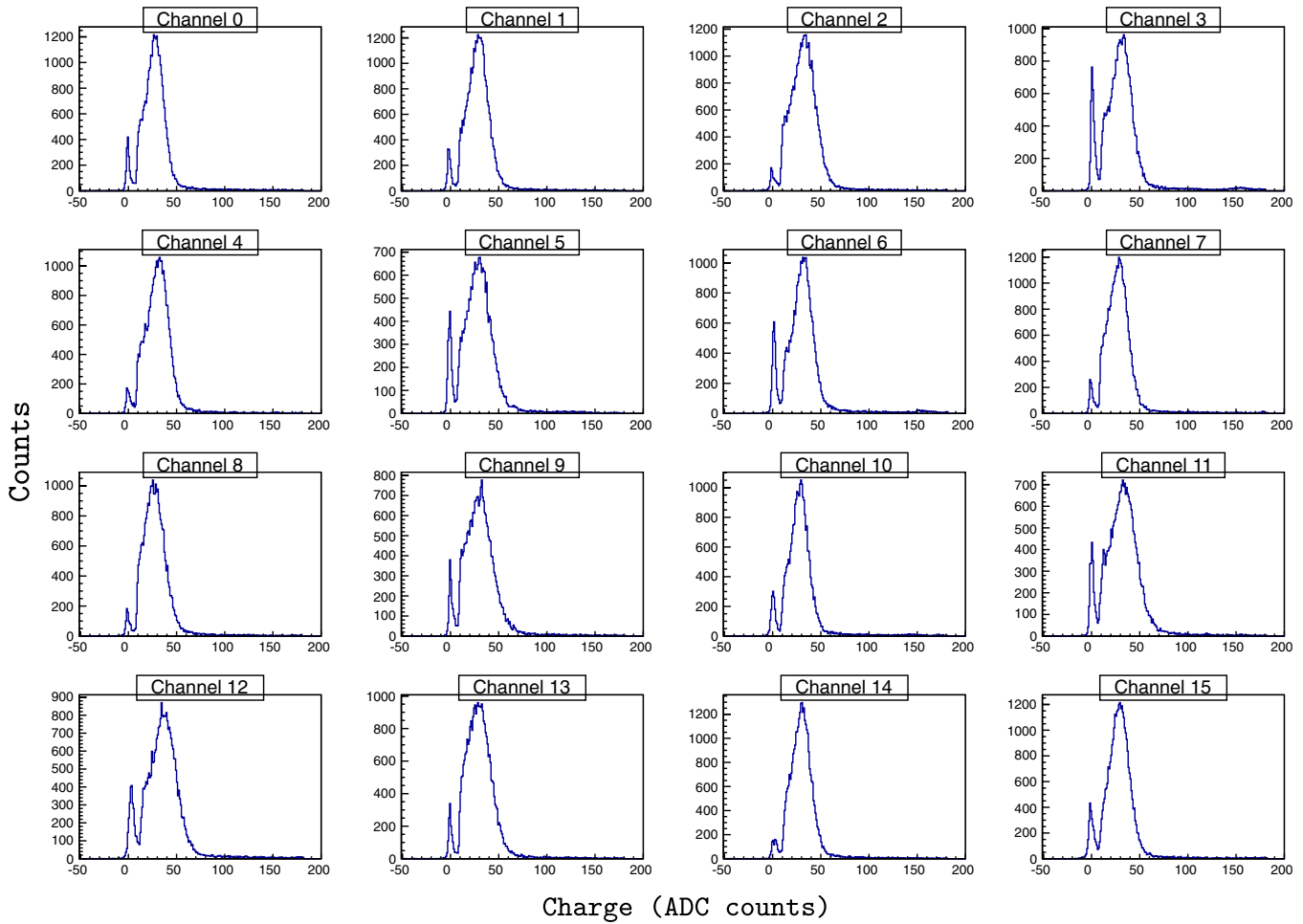

FIG. 3.   Recorded charge signals from cosmic muons in a $4 \times 4$ photomultiplier matrix equipped with grouped readout electronics and installed in the MEMPHYNO facility.

A schematic view and a photograph of the detector are shown in Fig. 2.

The $4 \times 4$ PMT matrix with the readout card developed by PMm$^2$ was installed in MEMPHYNO and signals from cosmic muons were recorded. Figure 3 shows the distributions of collected charge in the 16 PMTs with a large sample of muon events: they are consistent with the expectations from Cherenkov light emission, showing, above the pedestal, a peak with similar amplitude for all the channels. R&D is ongoing to characterize the response of the matrix and to optimize the performance.

## IV. MEMPHYS SIMULATION

In order to evaluate a realistic performance for the above-described baseline detector, a detailed simulation has been developed, mainly in the context of the EUROnu EU-FP7 Design Study [9]. The code, based on the GEANT-4 toolkit [22,23], was originally written for the T2K-2km detector [24], then interfaced with the OPENSCIENTIST framework [25]. It allows for interactive event viewing, batch processing and analysis. Special care has been devoted to the modularity of the code in the definition of the detector geometry, to facilitate future detector optimization studies. The GENIE [26] event generator is used for neutrino interactions.

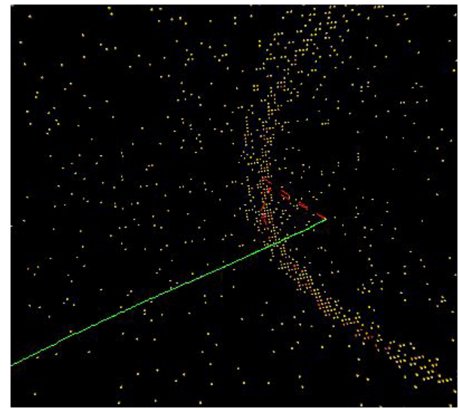

FIG. 4.   Pattern of hit PMTs after the interaction of a 500 MeV muon with the full MEMPHYS simulation. The green line is the muon track, the red dashed lines are gammas from muon capture, each white dot represents one hit PMT.

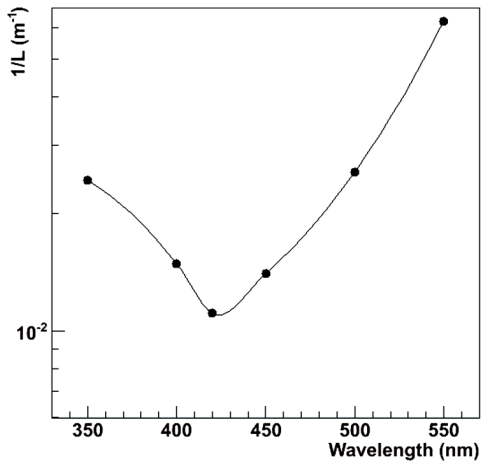

FIG. 5. Absorption coefficient for light in water as a function of wavelength in the GEANT-4 simulation of the MEMPHYS detector.

Figure 4 shows the pattern of hit PMTs following the interaction of a 500 MeV muon.

The correct modeling of light propagation in water has a crucial impact on the detector performance and must be faithfully reproduced by the simulation. The most relevant physical processes were implemented in GEANT-4: Rayleigh scattering, Compton scattering, and absorption. The attenuation length obtained as a function of wavelength is presented in Fig. 5. It reproduces well the one shown by SK in Fig. 13 of Ref. [27], thus providing evidence for the reliability of the simulation.

## V. STUDIES ON THE DETECTOR GEOMETRY

As mentioned above, the MEMPHYS design has evolved through different scenarios: three tanks with 65 m height, then three tanks of 80 m height, and finally two tanks of 103 m height. In order to ensure that the new design does not affect the detector performance, the response of MEMPHYS with different tank heights has been evaluated with the full simulation.

One basic quantity to consider is the number of detected photoelectrons (*PE*s) as a function of the particle energy. This is shown in Fig. 6 (top), for electrons generated uniformly in the detector volume, for three different tank heights. The number of *PE*s per MeV is about constant and equal to 11 for energies above 5 MeV. Figure 6 (bottom) shows the average number of hit PMTs as a function of energy for the three tank heights. The differences among the three configurations, while statistically significant, are not important for the physics that is proposed to be done with the detector.

We can conclude that the physics performance of the detector is independent of the tank height, in the range considered for the possible design. The new baseline configuration, with two tanks of 65 m diameter and 103 m height, is used in the following analysis.

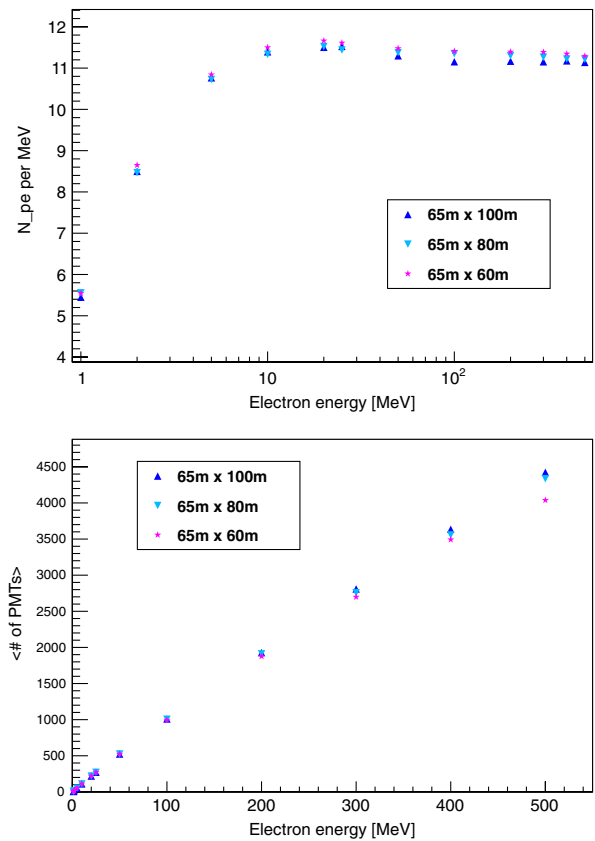

FIG. 6. Comparison of detector performance for three tank sizes, with 65, 80, and 100 m height. Top: Number of detected photoelectrons per MeV as a function of electron energy. Bottom: Average number of photomultipliers with at least one photoelectron, as a function of electron energy. The statistical error bars are smaller than the markers.

## VI. ANALYSIS ALGORITHMS

A complete analysis chain has been developed, based on what is done in SK. Some of the algorithms are simplified versions of those used by SK. Their performance was also evaluated by running the full simulation with the SK parameters (size, PMT coverage, etc.) to ensure that no significant degradation of efficiencies and background rejection are introduced by our algorithms.

The aim of the procedure is the reconstruction of the incoming neutrino energy and the identification of its flavor, to perform appearance or disappearance analyses with the different types of beams. This is only relevant for charged current (CC) neutrino interactions on nucleons ($N = p$ or $n$):

$$\nu_l + N \rightarrow l + N', \quad (1)$$

with $l = e$ or $\mu$ (the beams considered here are below threshold for $\tau$ production). Neutral current (NC) interactions

$$\nu_l + N \rightarrow \nu_l + X, \quad (2)$$

with X representing the generic hadronic final state, where a pion can mimic an electron or muon, are considered separately.

The analysis proceeds through the following steps: (i) reconstruction of the interaction vertex, from the timing of the hits in the different PMTs; (ii) determination of the outgoing lepton direction, from the pattern of the Cherenkov ring; (iii) lepton identification, from the "fuzziness" of the Cherenkov ring; (iv) rejection of NC interactions with a $\pi^0$ in the final state, from ring counting; and (v) reconstruction of the lepton momentum from the charge collected on the PMTs.

The incident neutrino energy is then deduced from the measured lepton momentum and direction.

Each step is described in detail in the following.

### A. Vertex and lepton direction reconstruction

The first step of the analysis is the reconstruction of the interaction vertex, by a fitting procedure which uses the timing information from all the hit PMTs. This provides a rough estimate of the vertex position of each event. The time residuals on the reconstructed vertex for electrons are shown in Fig. 7. The vertex position is identified by the point where the distribution of the time residuals of the PMTs has the sharpest peak. The estimator for a point source is taken from [28]:

$$G_p = \frac{1}{N} \sum_{i=1}^{n} \exp\left(-\frac{(t_i - t_m)^2}{2(1.5\sigma)^2}\right),$$

where $N$ is the number of hit PMTs, $t_i$ the time residual of the $i$th PMT, $t_m$ is the value where the time residuals distribution has its maximum, and $\sigma$ is the timing resolution of the PMTs (taken to be 2.5 ns). Assuming that all the photons are emitted from a point source at the same time, a coordinate grid scan over the detector volume is performed and the point where $G_p$ is maximized is chosen as the vertex. This algorithm works and is accurate for pointlike tracks (electrons, low energy muons) in MEMPHYS. The SK resolution figures [29] are applied to the true vertex point due to the lack of a precise vertex fitting algorithm for long tracks (e.g. muons); this is justified by the fact that, as will be shown later in this paper, the MEMPHYS reconstruction performance is very similar to that of SK.

The particle direction for single rings is reconstructed by taking into account the vertex point and the charge distribution inside MEMPHYS. A first estimate is calculated from the sum of all the unit vectors which begin at the reconstructed vertex point and end at each PMT, with each vector weighted by the charge in the PMT:

$$\vec{d}_0 = \sum_{i=1}^{} q_i \frac{\vec{P}_i - \vec{O}_0}{|\vec{P}_i - \vec{O}_0|},$$

where $\vec{d}_0$ is the particle direction, $\vec{O}_0$ is the vertex position obtained with the vertex fit described above, $\vec{P}_i$ is the position, and $q_i$ is the detected charge of the $i$th PMT. The distribution of detected charge as a function of the angular distance $\theta$ from the track direction, $PE(\theta)$, shown in Fig. 8, is considered and the following ring properties are calculated: (i) the angular distance from the track direction where the detected charge has its maximum, $\theta_{max}$; (ii) the angular distance from the track direction corresponding to the outer edge of the Cherenkov ring, $\theta_{edge}$, defined as the value of $\theta$, larger than $\theta_{max}$, for which the second derivative of the $PE$ distribution with respect to $\theta$ is zero.

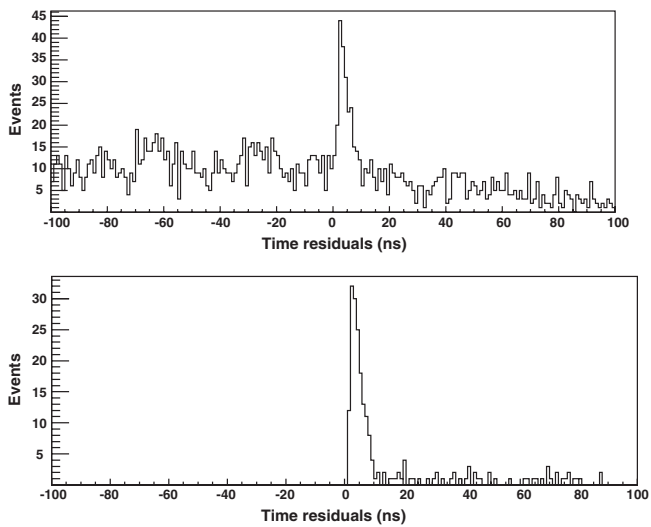

FIG. 7. Distribution of the time residuals, defined as the difference between the hit time measured at each PMT and the time of light propagation from the fitted vertex, for 25 MeV electrons. In the top plot, only the true vertex is considered. In the bottom plot, both the true vertex and the other points used for the grid scan (see text) are included.

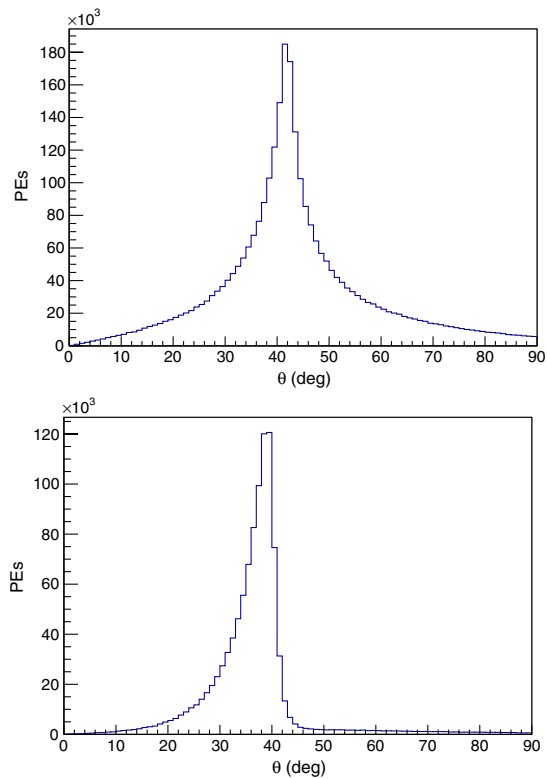

FIG. 8. Distribution of detected photoelectrons (PEs) as a function of the angular distance from the track direction ($\theta$) for 300 MeV electrons (top) and 300 MeV muons (bottom).

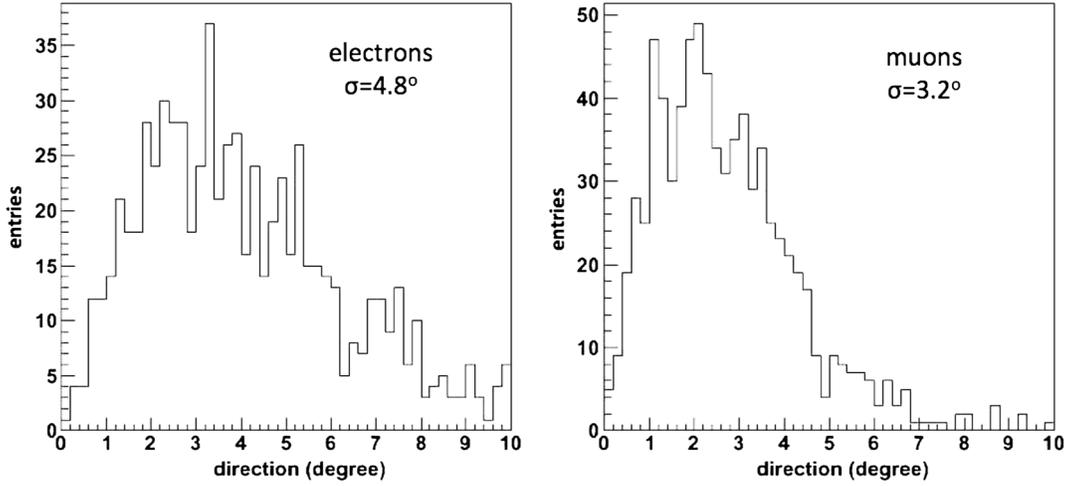

FIG. 9. Difference between the true and reconstructed direction for electrons (left) and muons (right) with energies uniformly distributed up to 1 GeV, in the full MEMPHYS simulation. The 68% integral is taken to define the resolution.

The precise direction is then reconstructed using the estimator

$$Q(\theta_{edge}) = \frac{\int_0^{\theta_{edge}} PE(\theta)d\theta}{\sin\theta_{edge}} \times \left(\frac{dPE(\theta)}{d\theta}\bigg|_{\theta=\theta_{edge}}\right)^2 \times \exp\left(-\frac{(\theta_{edge} - \theta_m)^2}{2}\right),$$

where $\theta_m$ indicates the mean value of the $PE(\theta)$ distribution. The estimator $Q(\theta_{edge})$ is calculated by changing the ring direction with respect to $\vec{d}_0$ into all possible directions within a cone with 20° half opening. The best particle direction is chosen as the one which maximizes $Q(\theta_{edge})$.

The resolution in lepton direction is shown in Fig. 9 for electron and muon samples and in Fig. 10 as a function of momentum for electrons and muons.

### B. Particle identification

Particle identification in MEMPHYS consists of (i) the separation of electrons from muons originating in CC interactions, to identify the flavor of the incoming neutrino; (ii) the recognition of $\pi^0$'s originating in NC interactions and decaying to two photons, which can mimic an electron and thus represent a background for the $\nu_e$-CC sample.

The two types of particle identification are described in this section.

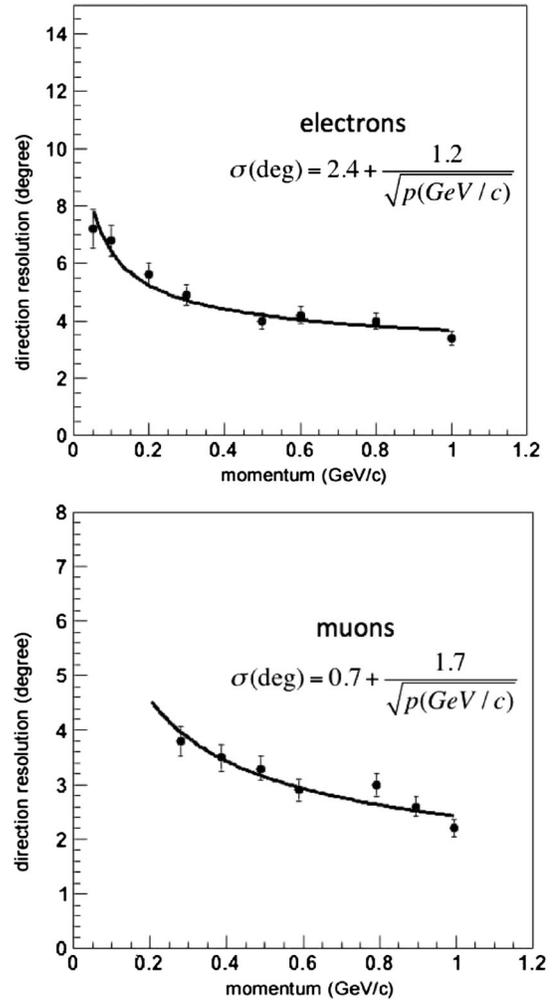

FIG. 10. Resolution in lepton direction as a function of momentum for electrons (top) and muons (bottom) in the full MEMPHYS simulation.

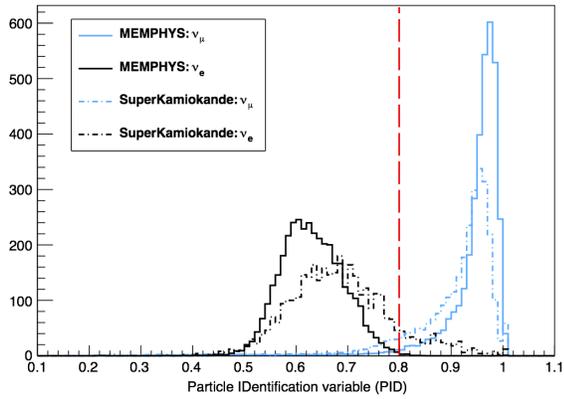

FIG. 11. Distributions of the particle identification (PID) variable for 500 MeV electrons and muons. The solid lines show the results with the MEMPHYS simulation, while the dashed ones are obtained when running the simulation with the geometry parameters of the SK detector and are presented for comparison. The vertical line indicates the cut value applied for identification.

The final state lepton is identified on the basis of the fuzziness of the Cherenkov ring, as in SK. Since electrons are subject to more multiple scattering and showering, the edges of the rings originating from them are less sharp than those of muon rings. Rings are thus classified as ''fuzzy'' (or e-like) or ''sharp'' ($\mu$-like).

The variable used for particle identification (PID) is the ratio of the charge contained inside the ring edge ($\theta_{\text{edge}}$, defined above) to the total charge in the event. This is simpler to compute that the likelihood function based on charge distributions, used by SK, relying on detailed comparisons of data and simulation, but proves to be very effective for electron/muon separation. Distributions of this PID quantity for single electron and single muon events are shown in Fig. 11. The distributions obtained when the simulation is run with the SK geometry parameters are also shown, for comparison; the edges of the Cherenkov cone are less sharp than in SK, due to more light scattering with the larger size of the detector, however the separation between the two types of leptons is better in MEMPHYS, mainly due to the larger number of PMTs providing better granularity when reconstructing the pattern of detected light. A cut at a value of 0.8 on this variable provides an efficiency of 96.6% for muons and 99.8% for electrons.

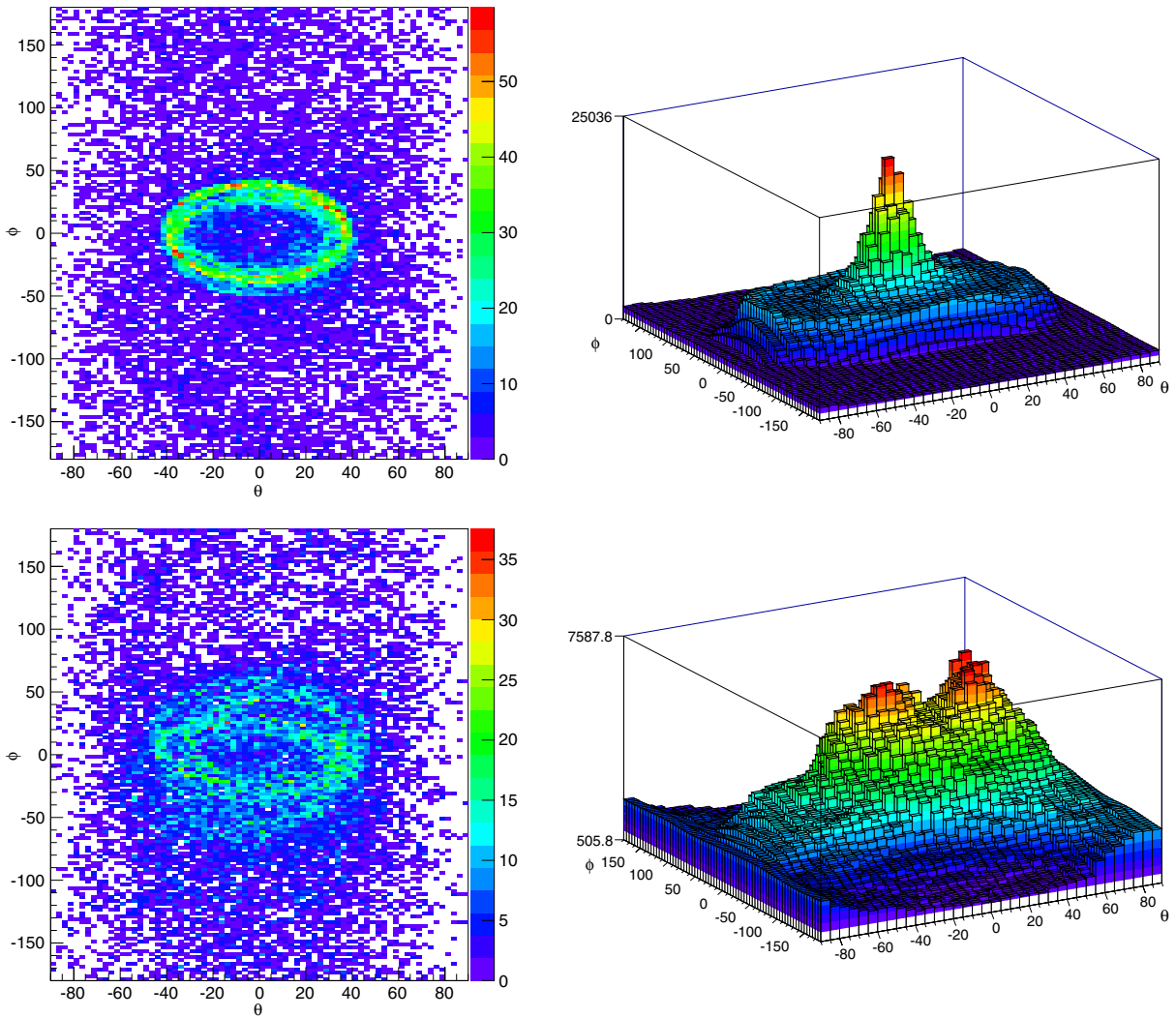

FIG. 12. Single-ring (top) and double-ring (bottom) events: projection in spherical coordinates centered on the reconstructed vertex and direction (left) and their Hough transform (right).

One of the most severe backgrounds in the search for $\nu_e$ appearance is due to NC events with a $\pi^0$ in the final state: the two $\gamma$'s originating from its decay produce rings similar to those of electrons, and the rejection of these events is mainly based on the reconstruction of a second ring in an electronlike event. The search for rings is performed with the method of the Hough transform [30], already used in SK (see for example [31]). This transform essentially consists in mapping each hit onto a circle centered on the hit PMT and weighted by its charge, and has the effect of transforming each ring into a peak. Peak counting is then performed, using two one-dimensional projections of the Hough-transformed distribution.

Figure 12 shows examples of a single-ring (electron) and a double-ring ($\pi^0$) event: the rings are first projected into spherical coordinates centered on the fitted particle vertex and direction, then Hough transformed to peaks. The $\pi^0$ identification algorithm used in this analysis is simpler than the one used in SK and in the HyperKamiokande LOI [32]: after rejecting events with more than one ring clearly identified, they force a second ring to be found in one-ring events and apply a cut on the invariant mass of the two photons around the $\pi^0$ mass. This particular cut was not applied in the present analysis, thus its efficiencies were rescaled to those of [32]: we can assume that we will eventually implement their full likelihood analysis and cuts and that the two detectors will perform rather closely on this particular aspect, as can be reasonably assumed from other studies presented above. We have applied our $\pi^0$ identification algorithm on a sample of neutrinos interacting in a detector simulated with an approximate SK geometry (40 m diameter, 40 m height, 40% optical coverage with 20" PMTs), and we have found a $\pi^0$ contamination very similar to what we find with the MEMPHYS simulation, namely, 3.9%: this suggests that we can assume the efficiency of the selection to be nearly independent of the detector size.

A cut on the Michel electron from muon decay was also introduced, since all the muons in the energy range of interest for the neutrino beams considered here (well below 2 GeV) are fully contained in the detector, and the detection of the decay electron is a powerful tool for their identification. The efficiencies for this cut as well were rescaled to those of [32]. The probability to find a Michel electron candidate (defined as at least 30 hits in a 50 ns window) after a $\nu_e$ interaction is negligible, therefore the $\nu_e$ contamination in the $\nu_\mu$ sample is completely suppressed. This cut introduces some differences between muon neutrino and antineutrino identification efficiency,

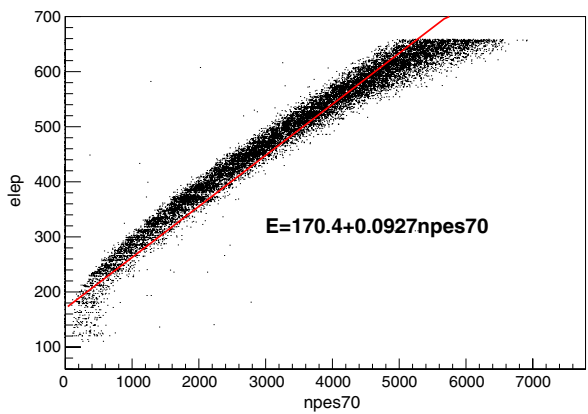

FIG. 13. Correlation between true and reconstructed momentum for electrons produced in $\nu_e$ CC interactions. The variable npes70 represents the number of photoelectrons detected in PMTs lying inside a cone with 70° half-opening angle, used as estimator of the lepton momentum.

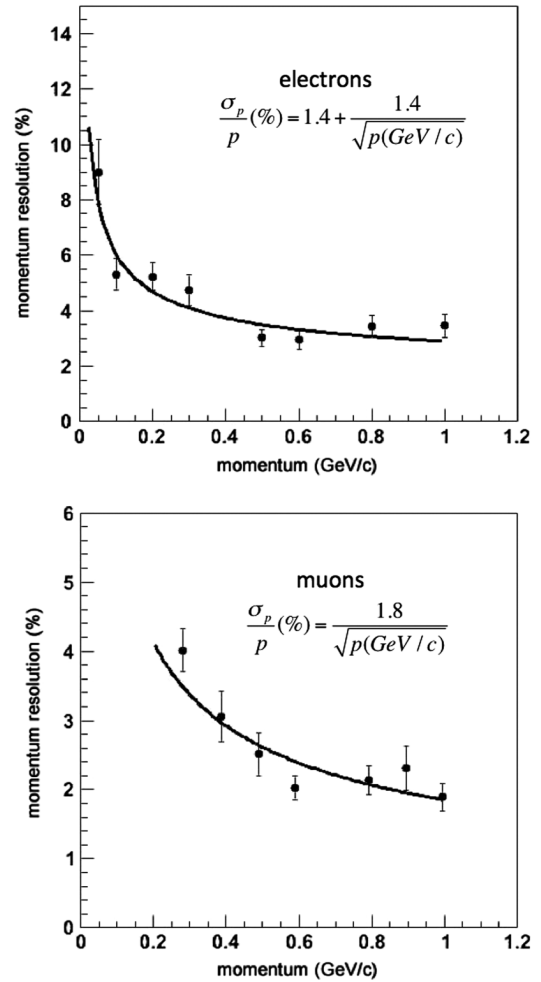

FIG. 14. Resolution on electron momentum (top) and muon momentum (bottom) in the full MEMPHYS simulation.

due to the different capture probabilities for $\mu^+$ and $\mu^-$ on nuclei.

### C. Lepton momentum determination

The momentum of the final state lepton can be reconstructed on the basis of the charge pattern in the PMTs. A simple estimator is used here: the observed *PE*s inside a cone with half-opening angle of 70 degrees are summed. This is found to show a simple almost-linear correlation with the lepton momentum, as shown in Fig. 13.

The resolution on the reconstructed lepton momentum is shown in Fig. 14. The MEMPHYS resolution is between those of SK-I and SK-II phases of the SK detector. This can be justified by simple considerations: at high energies, the resolution is determined essentially by the statistics of collected photoelectrons, which is the same as in SK with 20'' PMTs and 40% optical coverage; at lower energies, the larger detector size induces a small degradation.

## VII. NEUTRINO ENERGY RECONSTRUCTION

The incident neutrino energy is deduced from the measured lepton momentum and direction, assuming the interaction to be CC and quasielastic (QE). In a pure two-body collision such as (1), and assuming the nucleon is at rest, the incoming neutrino energy $E_\nu$ is related by simple kinematics to the outgoing lepton energy $E_l$ and momentum $P_l$ and to the angle $\theta_l$ of the lepton direction with respect to the neutrino:

$$E_\nu = \frac{m_N E_l - m_l^2/2}{m_N - E_l + P_l \cos\theta_l} \qquad (3)$$

with $m_N$ and $m_l$ denoting the nucleon and lepton masses. The direction of the incoming neutrino is that of the beam, which is precisely known, therefore the resolution on $\theta_l$ is entirely determined by the resolution on the lepton direction. The lepton momentum measurement and its resolution were described in Sec. VI C.

The reconstructed neutrino energy in two different energy ranges is shown in Fig. 15: the Gaussian peak is due to true QE interactions, with a smearing induced by the Fermi motion of the nucleon and the experimental resolution, while the shoulder at lower reconstructed energies is due to non-QE interactions, whose contribution is larger as energy increases.

### A. Migration matrices

In order to properly take into account all of the effects of the reconstruction, the detector performance is conventionally described in terms of "migration matrices" representing the neutrino reconstructed energy versus the true one. Each "slice" of true energy is normalized such that the projection of the matrix corresponds to the efficiency for the given neutrino energy. Separate matrices are con-structed for signal and background in the different detection channels, and for CC and NC events. They are computed for neutrinos, and consistent values were found for antineutrino interactions.

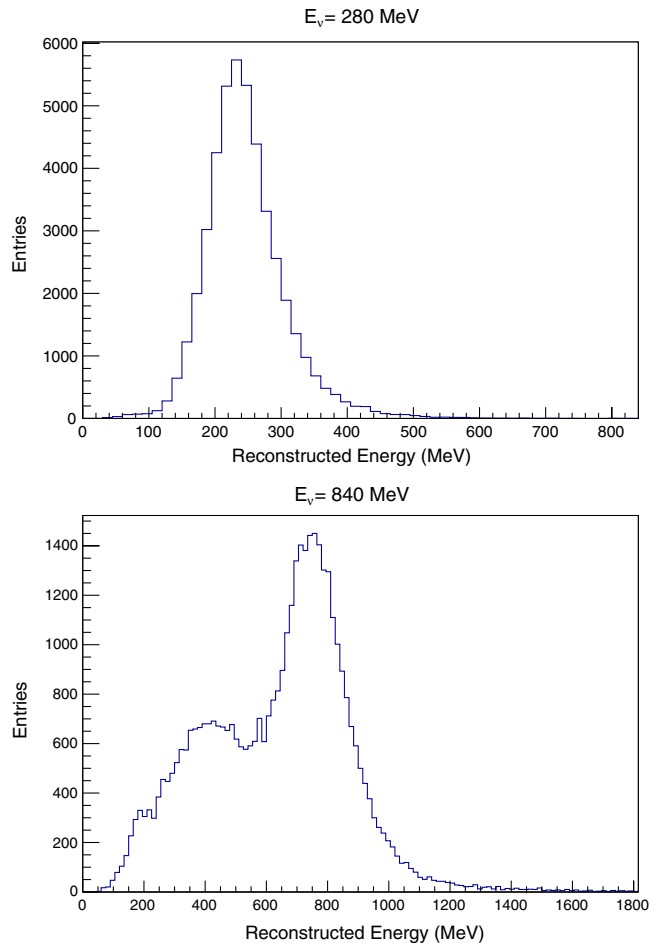

FIG. 15. Reconstructed energy for selected muon neutrinos with energies of 280 MeV (top) and 840 MeV (bottom).

Events identified as electron neutrinos represent the signal in the appearance channel in a Super-Beam, where the oscillation $\nu_\mu \to \nu_e$ is looked for. Separate migration matrices are provided for CC and NC interactions. The background is represented by misidentified $\nu_\mu$ CC interactions as well as by other components present in the beam in small fractions (mainly $\nu_e$'s and antineutrinos; no detailed study has been performed here for $\nu_\tau$'s, since the beam energy is below the threshold for $\tau$ production). Events identified as muons are the signal for the appearance channel $\nu_e \to \nu_\mu$ with a Beta-Beam or for the disappearance channel $\nu_\mu \to \nu_\mu$ with a Super-Beam.

The details of the matrices are provided in Fig. 16. The efficiencies as a function of neutrino energy are shown in Fig. 17.

The matrices [33] are available from the authors in the text format suitable as input for the GLOBES package [34,35].

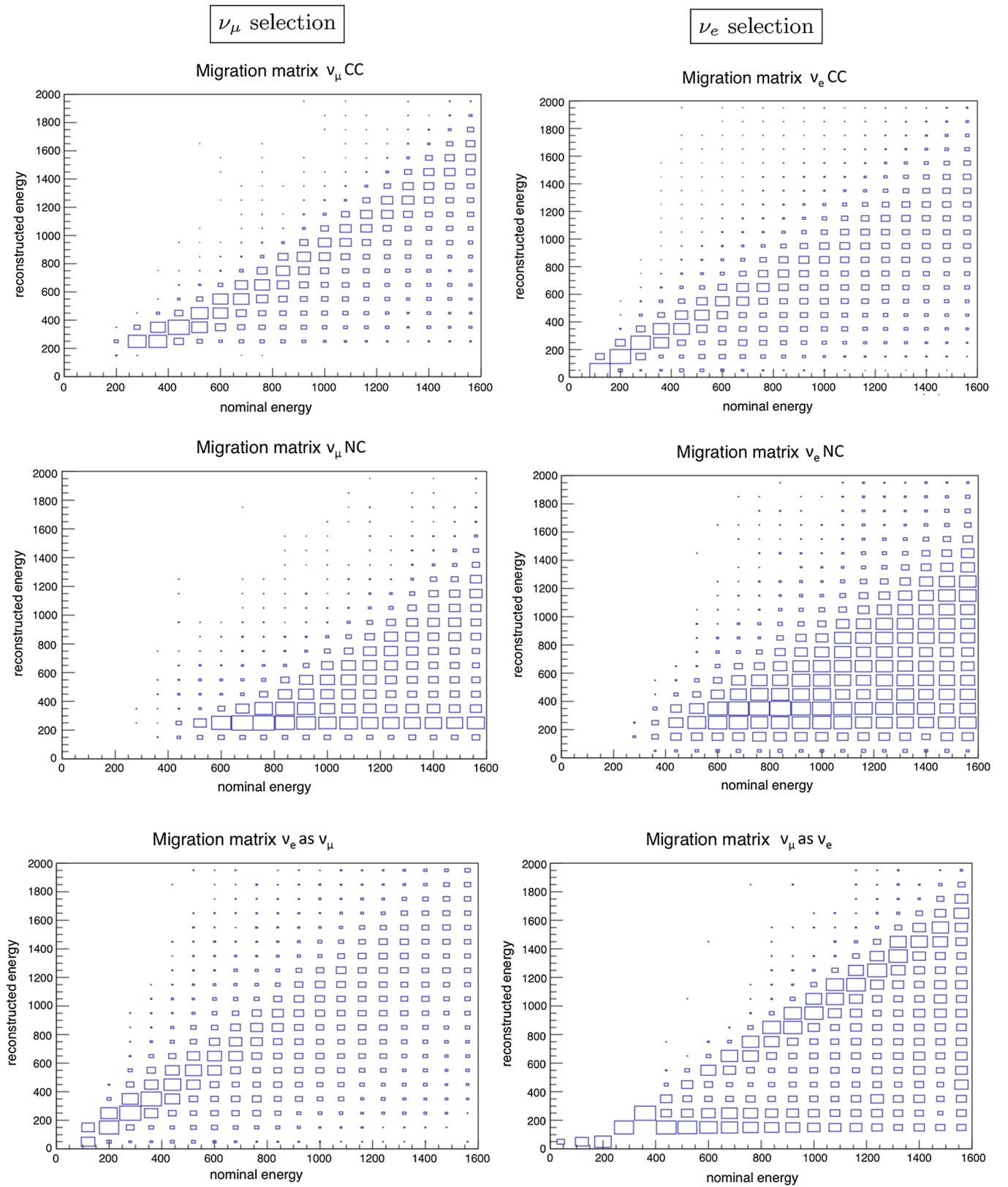

FIG. 16. "Migration matrices" with reconstructed neutrino energy as a function of true energy for selected events. Left: events identified as muon neutrinos, when they are $\nu_\mu$ CC interactions (top), NC interactions (middle), $\nu_e$ CC interactions (bottom). Right: events identified as electrons, when they are $\nu_e$ CC interactions (top), NC interactions (middle), $\nu_\mu$ CC interactions (bottom).

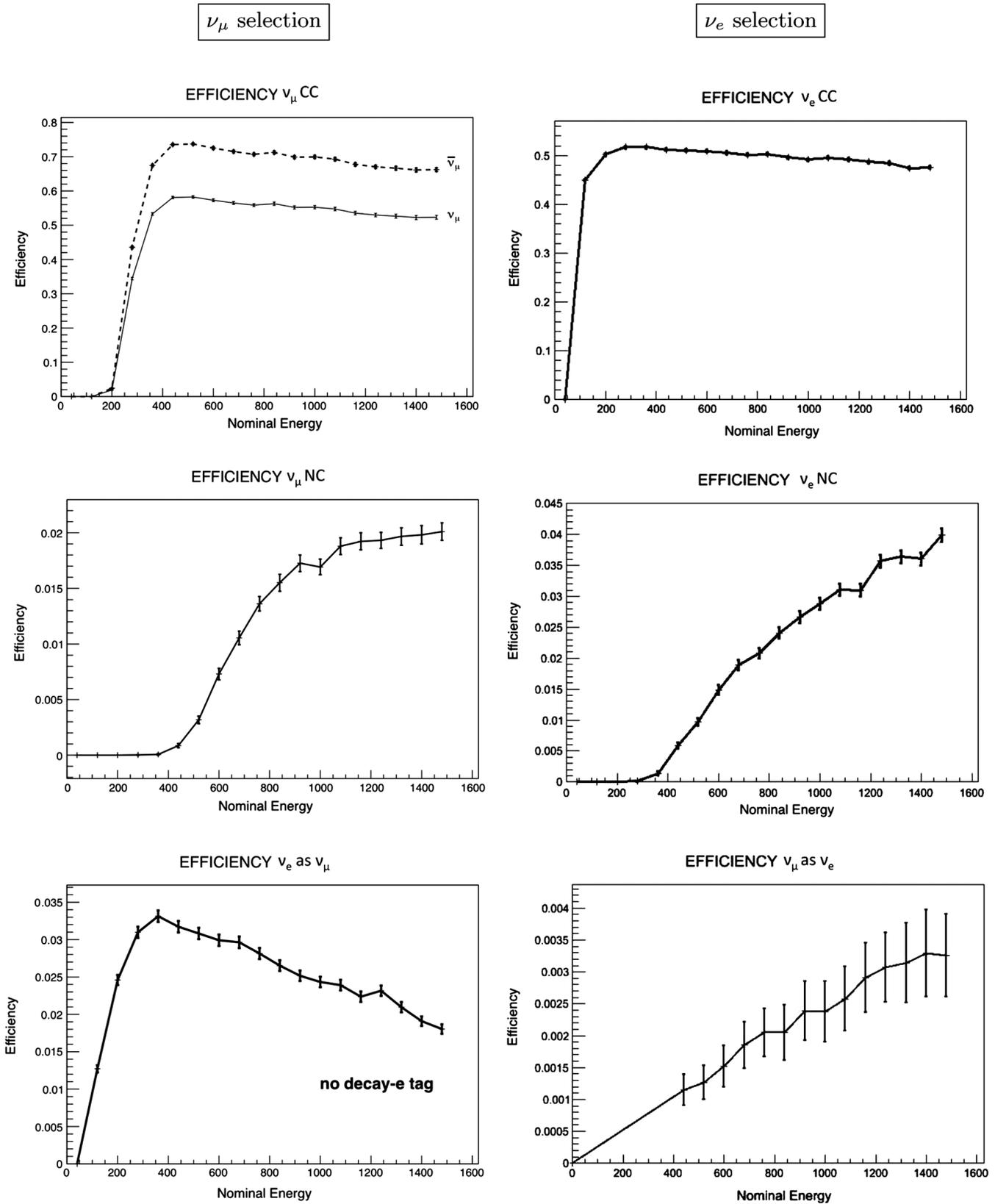

FIG. 17. Efficiencies for the selection of the different neutrino event categories in the MEMPHYS detector, as a function of neutrino energy. Left: events identified as muon neutrinos, when they are $\nu_\mu$ CC interactions (top), NC interactions (middle), $\nu_e$ CC interactions [(bottom). The cut decay-electron tag completely suppresses $\nu_e$ CC interactions and has not been applied for this plot.] Right: events identified as electrons, when they are $\nu_e$ CC interactions (top), NC interactions (middle), and $\nu_\mu$ CC interactions (bottom).

## VIII. SENSITIVITY STUDIES

Some examples of studies of physics reach using the GLOBES package are presented in this section. The option of the MEMPHYS detector at the Fréjus site with a neutrino beam from CERN is considered. Because of the relatively short baseline, oscillations are nearly unaffected by matter effects, therefore the experiment is well suited for LCPV but is nearly insensitive to mass hierarchy. Mass hierarchy can be measured with atmospheric neutrinos, as explained for example in [32], however this topic is beyond the scope of the present paper.

At a distance of 130 km, the first oscillation maximum occurs for energies close to 260 MeV. The two beams considered for these studies are particularly suited for this experimental setup: (i) a Beta-Beam [5] with an ion acceleration factor $\gamma = 100$, whose reference fluxes are $5.8 \times 10^{18}$ $^6$He useful decays/year and $2.2 \times 10^{18}$ $^{18}$Ne decays/year; (ii) a Super-Beam [10] with 4.5 GeV incident protons, providing $5.6 \times 10^{22}$ protons on target per year.

For the Beta-Beam, a running time of 5 years with neutrinos and 5 years with antineutrinos is considered, with a systematic uncertainty of 2% on both signal and background. For the Super-Beam, a running time of 2 years with neutrinos and 8 years with antineutrinos is considered, with a systematic uncertainty of 5% on signal and 10% on background. Normal mass hierarchy is assumed.

Figure 18 shows the sensitivity to the LCPV phase $\delta_{CP}$, as a fraction of values that can be distinguished from zero or $\pi$ at the given significance level ($3\sigma$ or $5\sigma$), as a function of the $\theta_{13}$ mixing angle. With the recently measured value of $\theta_{13}$, the experiment is sensitive to 65% of $\delta_{CP}$ values at $3\sigma$ with a Super-Beam and almost 80% with a Beta-Beam.

## IX. IMPACT OF SYSTEMATIC UNCERTAINTIES

Systematic errors in the search for neutrino oscillations with a Super-Beam are mainly related to the knowledge of the beam flux and composition, of neutrino interaction cross sections, and of the neutrino energy reconstruction in the detector.

A detailed description of expected systematic uncertainties with a neutrino experiment consisting of a Super-Beam and a megaton-scale water-Cherenkov detector is provided in [32]. The experience from analysis in the T2K experiment [36] is taken as a starting point and prospects for improvements are considered, concerning mainly (i) new hadron-production experiments which will improve our knowledge of neutrino fluxes, (ii) more data from the near detector, reducing systematics on neutrino cross sections, (iii) reduced uncertainty on antineutrino cross sections, if a magnetic field is used in the near detector, and (iv) larger statistics of atmospheric events at the far detector, to reduce detector-related uncertainties.

In the end, the authors of [32] estimate a 5% uncertainty on the signal and a 5% error on each of the remaining sources of systematic error: background from $\nu_\mu/\bar{\nu}_\mu$, background from $\nu_e/\bar{\nu}_e$, and relative normalization between neutrino and antineutrino. The latter contributions can be added in quadrature, yielding (with some conservative rounding) a systematic uncertainty on background of the order of 10%. This will be our baseline assumption for systematic errors. A very conservative estimate would be 15% for both signal and background systematic error, as in present experiments, while optimistic assumptions in past works [5,37] went as low as 2%.

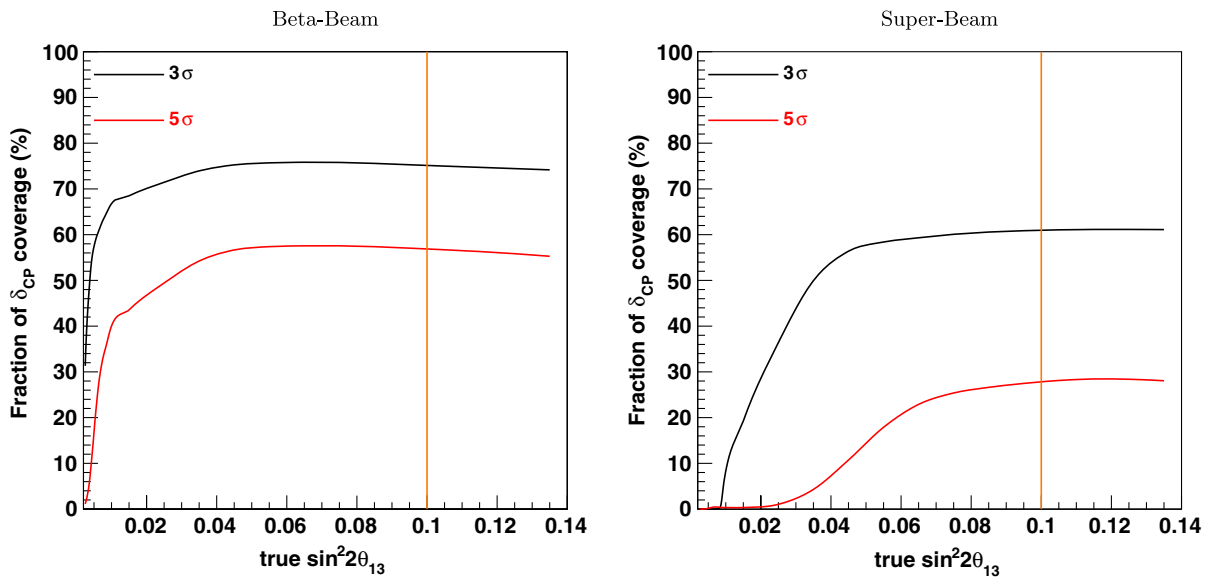

FIG. 18. Example of study of sensitivity to the leptonic *CP* violation phase using the GLOBES package, considering a Beta-Beam (left) or a Super-Beam (right) from CERN to the Fréjus site. The vertical line indicates approximately the currently measured value of the mixing parameter.

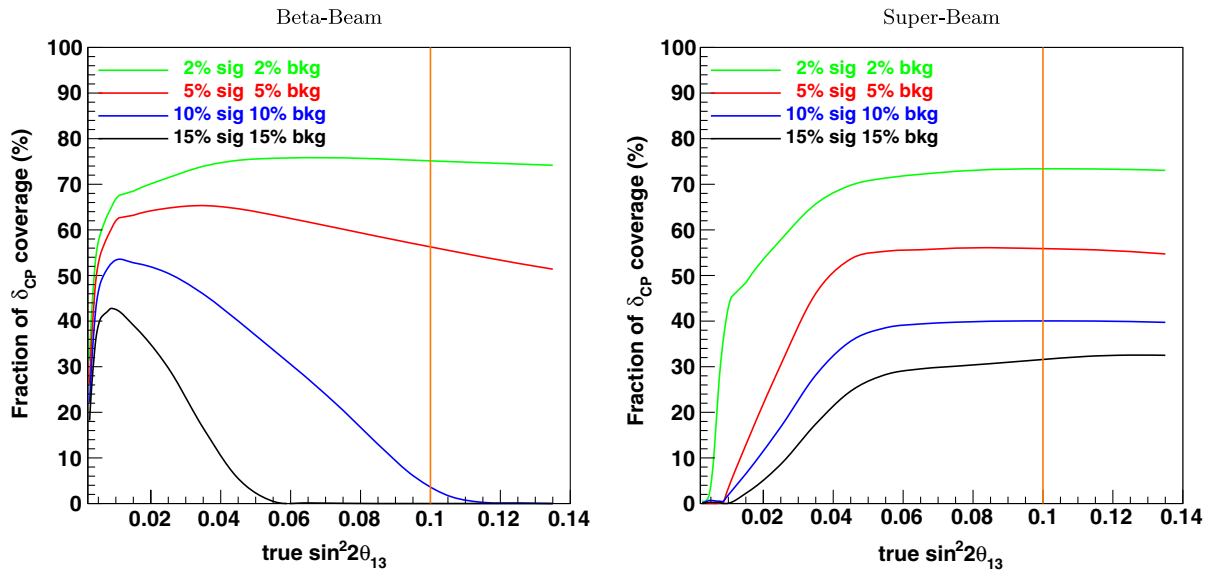

FIG. 19. Sensitivity to the leptonic *CP* violation phase, considering a Beta-Beam (left) and a Super-Beam (right) from CERN to the Fréjus site, with different assumptions on the systematic uncertainty on signal and background. The vertical line indicates approximately the currently measured value of the mixing parameter.

In the case of a Beta-Beam, the neutrino flux and spectrum are completely defined by the properties of the parent ion and its Lorentz boost. The main background is due to $\pi^{\pm}$'s generated from resonant processes in NC interactions: these pions cannot be separated from muons by particle identification. Contributions from $e/\mu$ misidentification and atmospheric events can be kept to a negligible level. The uncertainties on signal and background interaction cross sections at energies below 1 GeV are quite large, however the Beta-Beam itself can be the ideal place to measure them, if a near detector is installed. It has been estimated in [37] that a residual systematic error of 2% would be the final precision with which both the signal and the backgrounds can be evaluated. An additional effect will come from the uncertainties on the measurement of other oscillation parameters, but it will be very small.

The sensitivity to *CP* violation for the Beta-Beam and the Super-Beam experiments is shown in Fig. 19 with different assumptions on the size of the systematic uncertainty.

## X. CONCLUSIONS

A detailed study of the performance of a future large-scale water-Cherenkov detector, MEMPHYS, was performed, using a full simulation of the detector's response and realistic analysis algorithms. The sensitivity of a long baseline beam experiment was evaluated and the impact of systematic uncertainties was discussed. The MEMPHYS detector at the Fréjus site would have a $3\sigma$ sensitivity to about 65% of LCPV phase values with a Super-Beam from CERN and to 80% with a Beta-Beam. The baseline would be too short for a mass hierarchy measurement, which could however be performed with atmospheric neutrinos (not studied in this paper). R&D steps for the development of such a detector were also presented, focusing on electronics cards for grouped photomultiplier readout.

### ACKNOWLEDGMENTS

We are grateful to Enrique Fernandez Martinez for useful discussions and cross-checks and to Jaime Dawson for detailed reading of the manuscript. We acknowledge the financial support from the UnivEarthS Labex program of Sorbonne Paris Cité (ANR-10-LABX-0023 and ANR-11-IDEX-0005-02) and from the European Community under the European Commission Framework Programme 7 Design Study: EUROnu, Project No. 212372. The EC is not liable for any use that may be made of the information contained above.